\documentclass[a4paper]{IEEEtran}

\usepackage[utf8]{inputenc}  
\usepackage[T1]{fontenc}     
\usepackage[english]{babel}   

\usepackage{hyperref}		

\usepackage{amsmath}		
\usepackage{amssymb}		

\usepackage[b]{esvect}		

\usepackage{siunitx}		
\usepackage[version=4]{mhchem}		

\usepackage{graphicx}		
\usepackage{caption}		
\usepackage{subcaption}
	\newlength{\twosubht}	
	\newsavebox{\twosubbox}
\usepackage{tikz}			
	\usetikzlibrary{decorations.pathreplacing, calligraphy}	

\begin{document}
\title{A Fully Coupled Multi-Physics Model to Simulate Phase Change Memory Operations in Ge-rich \ce{Ge2Sb2Te5} Alloys}

\author{
\IEEEauthorblockN{R. Miquel$^{abc}$, T. Cabout$^{a}$, O. Cueto$^{b}$, B. Sklénard$^{b}$, M. Plapp$^{c}$}\vspace{0.2cm}\\
\IEEEauthorblockA{
$^{a}$ STMicroelectronics, 850 rue Jean Monnet 38926 Crolles, France \\
$^{b}$ Univ. Grenoble Alpes, CEA, Leti, F-38000 Grenoble, France\\
$^{c}$ Laboratoire de Physique de la Matière Condensée, Ecole Polytechnique, CNRS, Institut Polytechnique de Paris, \\91120 Palaiseau, France
}}

\maketitle

\begin{abstract}
A self-consistent model for the simulation of Ge-rich \ce{Ge2Sb2Te5} phase change memories is presented. Combining the multi-phase field model and a phase-aware electro-thermal solver, it reproduces the multi-physics behavior of the material. Simulations of memory operations are performed to demonstrate its ability to reproduce experimental observations.
\end{abstract}

\section{Introduction}
Phase change memories (PCM) relying on the electrical contrast between amorphous and crystalline phases of chalcogenide materials have been identified as a promising solution for embedded non-volatile memory technologies \cite{redaelli2022}. The Ge-rich \ce{Ge2Sb2Te5} (GGST) alloy has become a material of choice to satisfy the high-temperature retention requirement of the automotive market \cite{zuliani2013}.
While it provides excellent crystallization temperature and switching speed, this alloy exhibits complex phase change behaviors such as germanium segregation \cite{luong2021}. Predictive modeling of phase changes in the material is thus essential to aid technological development and to improve the understanding of underlying mechanisms.

In a previous work, a multi-phase field model (MPFM) has been proposed to simulate the evolution of GGST microstructure \cite{bayle2020}. However, the weak (one-way) coupling with an electro-thermal model and substantial approximations regarding phases properties inherent in this approach do not allow to fully describe memory operations. In this article, we report a complete phase change memory model obtained by a self-consistent coupling of the MPFM with an electro-thermal model. This enhanced coupling enables us to fully take into account the multi-physics aspect of PCM operations.

\section{Electro-thermal model}
\subsection{Equations}
The temperature $T$ in the structure is computed using the Fourier equation:
\begin{equation}
	C_p \frac{\partial T}{\partial t} = \vv{\nabla} . \left(k_{th}\vv{\nabla}T\right) + S
	\label{eq_fourier}
\end{equation}
with $k_{th}$ the thermal conductivity, $C_p$ the specific heat and $S$ the source term. The crystallization model distinguishes three phases: germanium (Ge), \ce{Ge2Sb2Te5} (GST) and amorphous/liquid (Am). This multi-phase aspect inherent to the coupling between MPFM and thermal model is reflected in physical parameters:
\begin{equation}
	k_{th}(T) = g_1 \, k_{th}^{Ge}(T) + g_2 \, k_{th}^{GST}(T) + g_3 \, k_{th}^{Am}(T)
	\label{eq_kth123}
\end{equation}
with $g_i$ weight coefficients associated with each phase, their sum being equal to 1 \cite{bayle2020}. The source term $S$ contains the contribution of latent heats induced by phase change and the Joule heating produced by the electric current:

\begin{equation}
	S = \underbrace{
	    - L_{Ge} \frac{\partial g_1}{\partial t}
	    - L_{GST}\frac{\partial g_2}{\partial t}
	    }_{\text{Latent heats of phase change}}
	    +  \underbrace{\sigma\big( \vv{\nabla}V \big)^2}_{\text{Joule heating}}
	\label{eq_source}
\end{equation}
with $L_{Ge}$ and $L_{GST}$ the latent heats of the two crystalline phases, $V$ the electrostatic potential and $\sigma$ the electrical conductivity.

In addition, considering the intensity of the Joule heating observed in PCM devices with respect to their small size, thermal boundary resistances at materials interfaces are taken into account.
Their introduction limits the heat flow at interfaces $\phi_i$, whose expression becomes $\phi_i = \Delta T/R_{i}$ with $\Delta T$ the temperature difference between both sides and $R_{i}$ the thermal resistance of the interface, leading to large temperature drops up to a few hundred kelvins. They are key to accurately reproduce the thermal confinement in the memory \cite{durai2020}.

Electrical conduction through the structure is responsible for both programming (via Joule heating) and reading operations (via resistivity sensing). It is computed using the Laplace equation. This approach, already used in PCM device simulations \cite{baldo2020}, is adapted to simulate Joule heating and electric current paths in the microstructure generated by the MPFM.
\begin{gather}
\vv{\nabla}.\vv{j} = 0 \quad \text{with} \quad \vv{j} = \sigma \vv{\nabla} V 	\label{eq_laplace} \\
\sigma(T, \vv{E}) = g_1 \, \sigma_{Ge}(T) + g_2 \, \sigma_{GST}(T) + g_3 \, \sigma_{\!Am}(T, \vv{E})			\label{eq_sigma(T)}
\end{gather}
with $\vv{j}$ the current density. The additional dependence of $\sigma_{\!Am}$ on the electric field $\vv{E}$ is important to reproduce the ovonic threshold switching effect that is key to PCM operations: at high electric field, the conductivity of the amorphous phase strongly increases to reach the crystalline one, enabling similar programming pulses regardless of the previous memory state \cite{guo2019}.

\subsection{Parameters}
The thermal conductivities of the three phases (Ge, GST and amorphous/liquid) in Eq. \eqref{eq_kth123} range over two orders of magnitude, resulting in different thermal behaviors in each phase.
For germanium, the temperature dependence from \cite{glassbrenner1964} is used as a starting point but the values are reduced to match the lower thermal conductivity in germanium thin films \cite{wang2011}.
For GST, measurements from \num{25} to \SI{400}{degreeCelsius} \cite{lyeo2006} are considered and extrapolated linearly up to the temperature where the melting of the phase occurs.
Finally, for the common Ge-rich GST amorphous/liquid phase, two dependences are considered:
one for the amorphous at low temperatures and one for the liquid at high temperatures.
They are connected linearly near the melting temperature.
A low thermal conductivity taken from \cite{kusiak2022} is considered for the amorphous.
For the liquid, no experimental data have been found.
As an alternative, DFT simulations values for liquid GST \cite{baratella2022} and measurements for liquid germanium \cite{assael2017} are then combined using the Filippov equation \cite{poling2001book}, resulting in a high thermal conductivity that increases with the temperature.

Latent heats used in \cite{bayle2020} are kept, and considered independent of the temperature. For the specific heat, the same constant value is used in the three phases.
In additional materials constituting the memory (see below), $k_{th}$ and $C_p$ are taken constant and material dependent.
Finally, the thermal boundary resistances are different for each pair of materials with values ranging over two orders of magnitude. 

Electrical conductivities were provided by unpublished in-house material characterization.
They increase with the temperature in the three phases of the model, but in the amorphous/liquid phase, several effects are also included.
The ovonic threshold switching effect is reproduced by using two temperature dependences for the amorphous, one at low electric field and one at high electric field.
At high temperature, they both increase rapidly to reach a higher electrical conductivity corresponding to the liquid \cite{crespi2014}.
Finally, a Poole-Frenkel conduction is also included in the low field dependence of the amorphous, increasing its conductivity for moderate electric fields.

\section{Implementation}
\subsection{Simulation domain}
\begin{figure}[h!]
\centering
\small
\begin{tikzpicture}
	\pgfdeclarelayer{bg}    
	\pgfsetlayers{bg,main}  

	\def\scale{0.016}		
	\def\lx{   300 * \scale};	
	\def\lxH{   10 * \scale};	
	\def\lxAc{ 150 * \scale};	
	\def\lyTE{ 100 * \scale};	
	\def\lyPCM{ 50 * \scale};	
	\def\lyMH{  75 * \scale};	
	\def\lyBE{  20 * \scale};	
	\def\ly{\lyTE + \lyPCM + \lyMH + \lyBE}	
	\def\lyBot{\lyMH + \lyBE}					
	\def\lxM{\lx/2 - \lxH/2};					
	\def\lxNAc{\lx/2 - \lxAc/2};				
	\def\xHD{\lx/2 + \lxH/2};	

	\draw (0, 0)           rectangle (\lx, \ly);
	\draw (0, \ly - \lyTE) --        (\lx, \ly - \lyTE);	
	\draw (0, \lyBot)      --        (\lx, \lyBot);		
	\draw (0, \lyBE)       --        (\lx, \lyBE);			
	\draw (\lxM,        \lyBE) -- (\lxM,        \lyBot);
	\draw (\lxM + \lxH, \lyBE) -- (\lxM + \lxH, \lyBot);
	\draw [dashed] (\lxNAc,         \lyBot) -- (\lxNAc,         \ly - \lyTE);
	\draw [dashed] (\lxNAc + \lxAc, \lyBot) -- (\lxNAc + \lxAc, \ly - \lyTE);

	\node at (\lx/2, \ly - \lyTE*.35) {Top electrode};
	\node at (\lx/2, \lyBot + \lyPCM/2) {GGST};
	\draw (\lx/2, \lyBE + \lyMH*0.6) -- (\lx*.55, \lyBE + \lyMH*0.85) node[right]{Heater};
	\node at (      \lx/4 - \lxH/4, \lyBE + \lyMH*.5) {Oxide};
	\node at (\lx - \lx/4 + \lxH/4, \lyBE + \lyMH*.4) {Oxide};
	\node at (\lx/2, \lyBE/2) {Bottom electrode};
	\draw [decorate, decoration = {brace, raise=2pt}] (\lxNAc, \ly-\lyTE) --  (\lxAc + \lxNAc, \ly-\lyTE);
	\node [above] at (\lx/2, \ly-\lyTE*0.92) {Active GGST};

	\begin{pgfonlayer}{bg}    	
		\path[fill=yellow, opacity=.4] (0,    \lyBE) rectangle (\lxM, \lyBot);
		\path[fill=red,    opacity=.5] (\lxM, \lyBE) rectangle (\xHD, \lyBot);
		\path[fill=yellow, opacity=.4] (\xHD, \lyBE) rectangle (\lx,  \lyBot);
		\path[fill=orange, opacity=.5] (0, 0)           rectangle (\lx, \lyBE);			
		\path[fill=blue,   opacity=.3] (0, \lyBot)      rectangle (\lx, \ly - \lyTE);	
		\path[fill=orange, opacity=.5] (0, \ly - \lyTE) rectangle (\lx, \ly);			
	\end{pgfonlayer}
	\end{tikzpicture}
\caption{Simulation domain.}
\label{fig_schema_memoire}
\end{figure}
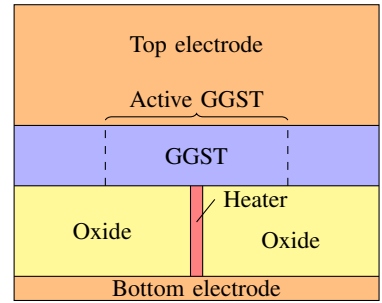

In PCM devices, efficient heating of the phase change material is obtained with a high and narrow conductor called heater, located beneath the GGST. An extended simulation domain of the whole memory cell (see Fig. \ref{fig_schema_memoire}) is thus needed to properly simulate the temperature field.
A total domain of \SI{300}{\nano\meter} width by \SI{240}{\nano\meter} high, in 2D, is used as a good trade-off between resolution speed and realistic thermal profile.

\begin{figure}[h!]
	\centering
	\small
	\begin{tikzpicture}
	\definecolor{colEltd}{rgb}{1   , .745, .510}	
	\definecolor{colGGST}{rgb}{.698, .698, 1   }	
	\definecolor{colOx}{rgb}{1   , .984, .604}	
	\definecolor{colH}{rgb}{1   , .510, .510}	
	\def\op{0.2}	
	\definecolor{CL}{rgb}{.0, .65, .0}

	\pgfdeclarelayer{bg}    
	\pgfsetlayers{bg,main}  

	\def\scale{0.0115}		
	\def\lx{   240 * \scale};	
	\def\lxH{   10 * \scale};	
	\def\lxAc{ 120 * \scale};	
	\def\lyTE{ 100 * \scale};	
	\def\lyPCM{ 50 * \scale};	
	\def\lyMH{  75 * \scale};	
	\def\lyBE{  20 * \scale};	
	\def\ly{\lyTE + \lyPCM + \lyMH + \lyBE}	
	\def\lyBot{\lyMH + \lyBE}					
	\def\lxM{\lx/2 - \lxH/2};					
	\def\lxNAc{\lx/2 - \lxAc/2};				
	\def\xHD{\lx/2 + \lxH/2};	
	\def\dyTitre{0.5}

	\node [below] at (\lx/2, \ly + \dyTitre) {Crystallization}; 
	\draw (0, 0)           rectangle (\lx, \ly);
	\draw (0, \ly - \lyTE) --        (\lx, \ly - \lyTE);	
	\draw (0, \lyBot)      --        (\lx, \lyBot);		
	\draw (0, \lyBE)       --        (\lx, \lyBE);			
	\draw (\lxM,        \lyBE) -- (\lxM,        \lyBot);
	\draw (\lxM + \lxH, \lyBE) -- (\lxM + \lxH, \lyBot);
	\draw [color=CL, line width=0.1cm](\lxNAc,         \lyBot) -- (\lxNAc,         \ly - \lyTE);
	\draw [color=CL, line width=0.1cm](\lxNAc + \lxAc, \lyBot) -- (\lxNAc + \lxAc, \ly - \lyTE);
	\draw [dashed] (\lxNAc,         \lyBot) -- (\lxNAc,         \ly - \lyTE);
	\draw [dashed] (\lxNAc + \lxAc, \lyBot) -- (\lxNAc + \lxAc, \ly - \lyTE);

	\begin{pgfonlayer}{bg}    	
		\path[fill=colOx, opacity=\op] (0,    \lyBE) rectangle (\lxM, \lyBot);
		\path[fill=colH,  opacity=\op] (\lxM, \lyBE) rectangle (\xHD, \lyBot);
		\path[fill=colOx, opacity=\op] (\xHD, \lyBE) rectangle (\lx,  \lyBot);
		\path[fill=colEltd, opacity=\op] (0, 0)           rectangle (\lx, \lyBE);			
		\path[fill=colEltd, opacity=\op] (0, \ly - \lyTE) rectangle (\lx, \ly);			
		\path[fill=colGGST, opacity=\op] (0,              \lyBot) rectangle (\lxNAc,         \ly - \lyTE); 
		\path[fill=colGGST]              (\lxNAc,         \lyBot) rectangle (\lxNAc + \lxAc, \ly - \lyTE); 
		\path[fill=colGGST, opacity=\op] (\lxNAc + \lxAc, \lyBot) rectangle (\lx,            \ly - \lyTE); 
	\end{pgfonlayer}

	\def\dx{3};
	\node [below] at (\lx/2 + \dx, \ly + \dyTitre + 0.04) {Thermal};
	\draw (\dx, 0)           rectangle (\dx + \lx, \ly);
	\draw (\dx, \ly - \lyTE) --        (\dx + \lx, \ly - \lyTE);	
	\draw (\dx, \lyBot)      --        (\dx + \lx, \lyBot);		
	\draw (\dx, \lyBE)       --        (\dx + \lx, \lyBE);			
	\draw (\dx + \lxM,        \lyBE) -- (\dx+\lxM,        \lyBot);
	\draw (\dx + \lxM + \lxH, \lyBE) -- (\dx+\lxM + \lxH, \lyBot);
	\draw [dashed] (\dx + \lxNAc,         \lyBot) -- (\dx + \lxNAc,         \ly - \lyTE);
	\draw [dashed] (\dx + \lxNAc + \lxAc, \lyBot) -- (\dx + \lxNAc + \lxAc, \ly - \lyTE);

	\begin{pgfonlayer}{bg}    	
		\path[fill=colOx] (\dx,        \lyBE) rectangle (\dx + \lxM, \lyBot);
		\path[fill=colH]  (\dx + \lxM, \lyBE) rectangle (\dx + \xHD, \lyBot);
		\path[fill=colOx] (\dx + \xHD, \lyBE) rectangle (\dx + \lx,  \lyBot);
		\path[fill=colEltd] (\dx, 0)           rectangle (\dx + \lx, \lyBE);			
		\path[fill=colGGST] (\dx, \lyBot)      rectangle (\dx + \lx, \ly - \lyTE);	
		\path[fill=colEltd] (\dx, \ly - \lyTE) rectangle (\dx + \lx, \ly);			
	\end{pgfonlayer}

	\draw [color=CL, line width=0.09cm](\dx, 0) rectangle (\dx + \lx, \ly);

	\def\dx{6};
	\node [below] at (\lx/2 + \dx, \ly + \dyTitre) {Electrical};
	\draw (\dx, 0)           rectangle (\dx + \lx, \ly);
	\draw (\dx, \ly - \lyTE) --        (\dx + \lx, \ly - \lyTE);	
	\draw (\dx, \lyBot)      --        (\dx + \lx, \lyBot);		
	\draw (\dx, \lyBE)       --        (\dx + \lx, \lyBE);			
	\draw (\dx + \lxM,        \lyBE) -- (\dx+\lxM,        \lyBot);
	\draw (\dx + \lxM + \lxH, \lyBE) -- (\dx+\lxM + \lxH, \lyBot);
	\draw [dashed] (\dx + \lxNAc,         \lyBot) -- (\dx + \lxNAc,         \ly - \lyTE);
	\draw [dashed] (\dx + \lxNAc + \lxAc, \lyBot) -- (\dx + \lxNAc + \lxAc, \ly - \lyTE);

	\begin{pgfonlayer}{bg}    	
		\path[fill=colOx, opacity=\op] (\dx,        \lyBE) rectangle (\dx + \lxM, \lyBot);
		\path[fill=colH]               (\dx + \lxM, \lyBE) rectangle (\dx + \xHD, \lyBot);
		\path[fill=colOx, opacity=\op] (\dx + \xHD, \lyBE) rectangle (\dx + \lx,  \lyBot);
		\path[fill=colEltd, opacity=\op] (\dx, 0)           rectangle (\dx + \lx, \lyBE);			
		\path[fill=colEltd, opacity=\op] (\dx, \ly - \lyTE) rectangle (\dx + \lx, \ly);			
		\path[fill=colGGST, opacity=\op] (\dx,                  \lyBot) rectangle (\dx + \lxNAc,         \ly - \lyTE); 
		\path[fill=colGGST]              (\dx + \lxNAc,         \lyBot) rectangle (\dx + \lxNAc + \lxAc, \ly - \lyTE); 
		\path[fill=colGGST, opacity=\op] (\dx + \lxNAc + \lxAc, \lyBot) rectangle (\dx + \lx,            \ly - \lyTE); 
	\end{pgfonlayer}

	\draw [color=CL, line width=0.09cm](\dx + \lxNAc,          \ly - \lyTE)
	                                -- (\dx + \lxNAc + \lxAc,  \ly - \lyTE); 
	\draw [color=CL, line width=0.09cm](\dx + \lxM - 0.05,         \lyBE)
									 -- (\dx + \lxM + \lxH + 0.05, \lyBE); 
	\end{tikzpicture}
    \caption{Resolution domains and boundary conditions associated with each model.  Resolution domains are represented using plain colors and Dirichlet boundary conditions are in green.}
    \label{fig_domain_model}
\end{figure}
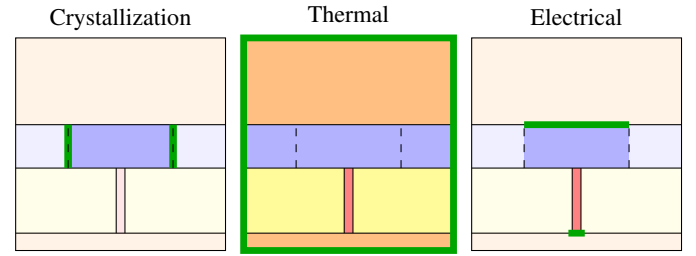

The thermal equations are solved on the full domain and the remaining equations are solved on specific areas (see Fig. \ref{fig_domain_model}). The MPFM is solved in the "active GGST" region, the only area where the temperature rise is sufficient to trigger a phase change. Its width has been chosen according to experimental observations. Similarly, the electrical model is solved on a reduced domain comprised of the GGST layer and the heater.
Boundary conditions also vary from one model to another. For the crystallization model, the continuity of the microstructure between active and non-active parts of the GGST region is maintained.
For the temperature, Dirichlet boundary conditions are applied all around the domain and fixed at \SI{300}{\kelvin}. Finally, for the electrical model, two potentials are applied on the top and the bottom electrodes (they can be adjusted to force a fixed current during the simulation), and zero-flux conditions are imposed at oxide interfaces and other domain boundaries.
A summary of the simulation domains and boundary conditions associated with each model is presented in Fig. \ref{fig_domain_model}.

\subsection{Numerical details}
The simulator is an in-house C++ implementation using finite differences and an explicit (forward Euler) time scheme.
The addition of the non-stationary thermal equation introduces a faster diffusive phenomenon that, for numerical stability reason, strongly decreases the simulation timestep $dt$
($dt \leq dx^2/4D$ with $D$ the largest diffusion coefficient and $dx$ the grid discretization).
This leads to prohibitive simulation time.
Increasing $dx$ is not an option because the phase field model requires multiple points in interfaces.
Instead, to overcome this issue, the thermal equation is computed on a coarser grid with $dx' = N.dx$ and is solved multiple times each simulation step to reach MPFM's $dt$.
We tuned $dx'$ and the number of times the equation is solved in each step to ensure an acceptable error on the results.
In addition, OpenMP parallelization is used to reduce the simulation time.

Equation \eqref{eq_laplace} being stationary, the electrical model is not affected by stability considerations. However, it is solved on the same thermal grid (larger $dx$) for performance reasons, using the conjugate gradient method.

\subsection{Thermal conductivity implementation}
Thermal conductivities are computed on the MPFM grid (smaller $dx$) while they are used on the thermal grid to solve Eq. \eqref{eq_fourier}.
As a consequence, a transfer between the two grids is needed.
To preserve physical integrity, the underlying thermal conduction is considered.
The heat going from one node to the next on the thermal grid actually passes through $N$ links (between $N+1$ nodes) on the MPFM grid.
The effective thermal conductivity $k_\text{eff}$ between the two thermal nodes therefore corresponds to those $N$ conductivities "in series" :
\begin{equation}
\frac{1}{k_\text{eff}} = \dfrac{1}{N}\sum\limits_{i=0}^{N-1} \dfrac{1}{k_{i,i+1}}
\label{eq_keff_MG}
\end{equation}
with $k_{i,i+1}$ the thermal conductivity between nodes $i$ and $i+1$ on MPFM grid.

\section{Simulation results} 
During the fabrication process of PCM, the GGST is fully crystallized.
Therefore, a layer made of crystalline germanium and GST is used as an initial condition for memory operations simulations (see Fig. \ref{fig_reset_structure}.1).
This layer is obtained by simulating the annealing of an amorphous domain in which germanium and GST grains nucleate and then grow using the MPFM model (see \cite{bayle2020}).
During the nucleation process, crystalline seeds are placed randomly, such that different runs lead to different grain distributions.

\subsection{RESET operation}
The evolution of the microstructure during a RESET operation simulation (pulse pattern of Fig. \ref{fig_pulses}) is presented in Fig. \ref{fig_reset_structure}.
From the initial fully crystallized layer, a dome-shaped domain (in black) is first melted thanks to the temperature profile presented in Fig. \ref{fig_reset_temp}, then quenched by the fast decrease of imposed current.
Between steps 3 and 4, a slight recrystallization begins but quickly stops as the temperature drops.

\begin{figure}[!htbp]
	\centering
	\includegraphics[width=0.9\columnwidth]{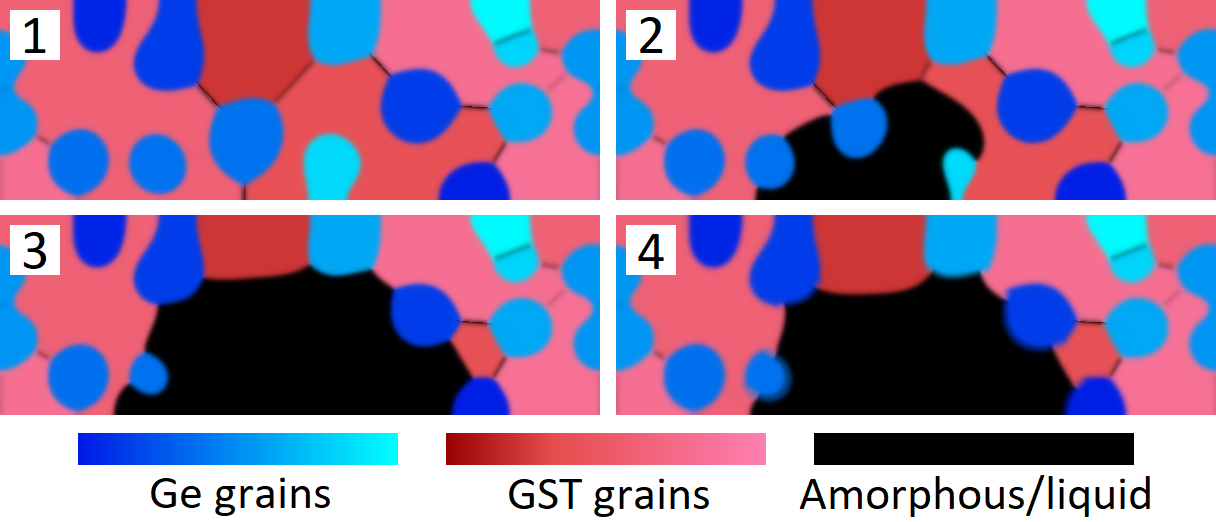}
	\caption{Microstructure evolution during a RESET pulse: before melting (1), during the pulse (2, 3), after operation (4). Germanium domains are blue, GST domains are red, amorphous and liquid domains are black. Different shades of blue and red mean different grain orientations (managed by the MPFM).}
	\label{fig_reset_structure}
\end{figure}
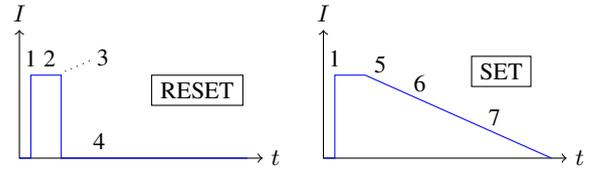
\begin{figure}[!htbp]
	\centering
	\small
	\begin{tikzpicture}
		\def\tPulse{0.4}		
		\def\vPulse{1.1}		
		\def\tDeb{0.15}
		\def\yReset{0}
		\def\xS{4}			
		\def\ySet{0}
		\def\tEnd{3}

		\foreach \x in {0, \xS} {
			\draw [<->] (\x, 1.7) node [above] {$I$} -- (\x, 0) --
			            (\x + \tEnd + .2, 0) node [right] {$t$};
		}

		\draw [blue] (0, \yReset) -- (\tDeb, \yReset) -- (\tDeb, \yReset + \vPulse) --
		             (\tDeb + \tPulse, \yReset + \vPulse) -- (\tDeb+\tPulse, \yReset) -- (\tEnd, \yReset);
		\draw [blue] (\xS, \ySet) -- (\xS + \tDeb, \ySet) --
		             (\xS + \tDeb, \ySet + \vPulse) --
		             (\xS + \tDeb + \tPulse, \ySet + \vPulse) -- (\xS + \tEnd, \ySet);

		\node [above] at (\tDeb, \yReset + \vPulse) {1};
		\node [above] at (\tDeb + \tPulse/2 + 0.05, \yReset + \vPulse) {2};
		\draw [dotted] (\tDeb + \tPulse + 0.05, \yReset + \vPulse + 0.05) --
		               (\tDeb + \tPulse + 0.4,  \yReset + \vPulse + 0.2);
		\node at       (\tDeb + \tPulse + 0.55, \yReset + \vPulse + 0.23) {3};
		\node [above] at (\tDeb + \tPulse + 0.5, \yReset) {4};
		\node [above] at (\xS + \tDeb, \ySet + \vPulse) {1};
		\node [above] at (\xS + \tDeb + \tPulse + 0.2 , \ySet + \vPulse - 0.08) {5};
		\node [above] at (\xS + \tDeb + \tPulse + 0.72, \ySet + \vPulse - 0.33) {6};
		\node [above] at (\xS + \tDeb + \tPulse + 1.7,  \ySet + \vPulse/2 - 0.23){7};

		\draw (\xS + \tEnd * 0.65, \ySet + \vPulse - 0.15) rectangle
		      (\xS + \tEnd * 0.91, \ySet + \vPulse + 0.25) node[pos=.5] {SET};
		\draw (\tEnd * 0.58, \yReset + \vPulse - 0.4) rectangle
		      (\tEnd * 0.98, \yReset + \vPulse) node[pos=.5] {RESET};
	\end{tikzpicture}
    \caption{RESET and SET electrical pulses used for simulations. The numbers correspond to the labels of the subfigures in Fig. \ref{fig_reset_structure}, \ref{fig_reset_temp_current} and \ref{fig_set_structure}.}
    \label{fig_pulses}
\end{figure}
\begin{figure}[h!]
	\sbox\twosubbox{%
		\resizebox{\dimexpr.49\textwidth-1em}{!}{%
			\includegraphics[height=.2cm]{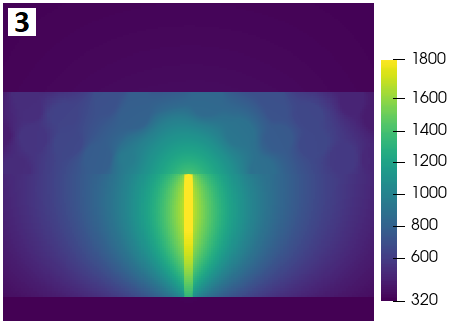}%
			\includegraphics[height=.2cm]{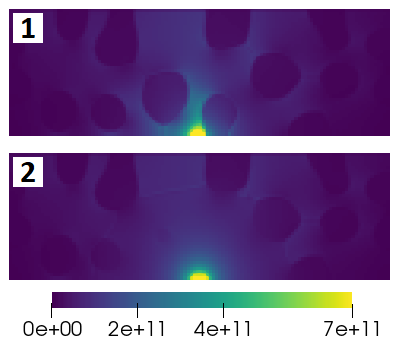}%
		}%
	}
	\setlength{\twosubht}{\ht\twosubbox}
	\centering
	\subcaptionbox{Temperature (K)\label{fig_reset_temp}}{%
		\includegraphics[height=\twosubht]{temp.png}%
	}\;
	\subcaptionbox{Current density (\si{\ampere\per\meter\squared}) \label{fig_reset_current}}{%
		\includegraphics[height=\twosubht]{current_RESET.png}%
	}
	\caption{Temperature and electric current density during RESET operation of Fig. \ref{fig_reset_structure}. (a) shows a zoom on the heater and (b) only shows the "active GGST" region.}
	\label{fig_reset_temp_current}
\end{figure}

Due to the different electro-thermal properties of the three phases, the non-uniformity of the GGST layer is clearly seen in temperature and current density maps presented in Fig. \ref{fig_reset_temp_current}. In particular, the higher melting temperature of germanium grains is not compensated by an enhanced Joule heating due to their high electrical resistivity. Instead, the electric current avoids those grains and their melting is mainly attributed to the heat accumulation in the liquid dome, leading to temperatures high enough to melt germanium. The final structure is consistent with experimental measurements \cite{sousa2015}.

\subsection{SET operation}
\begin{figure}[!h]
	\centering
	\includegraphics[width=0.9\columnwidth]{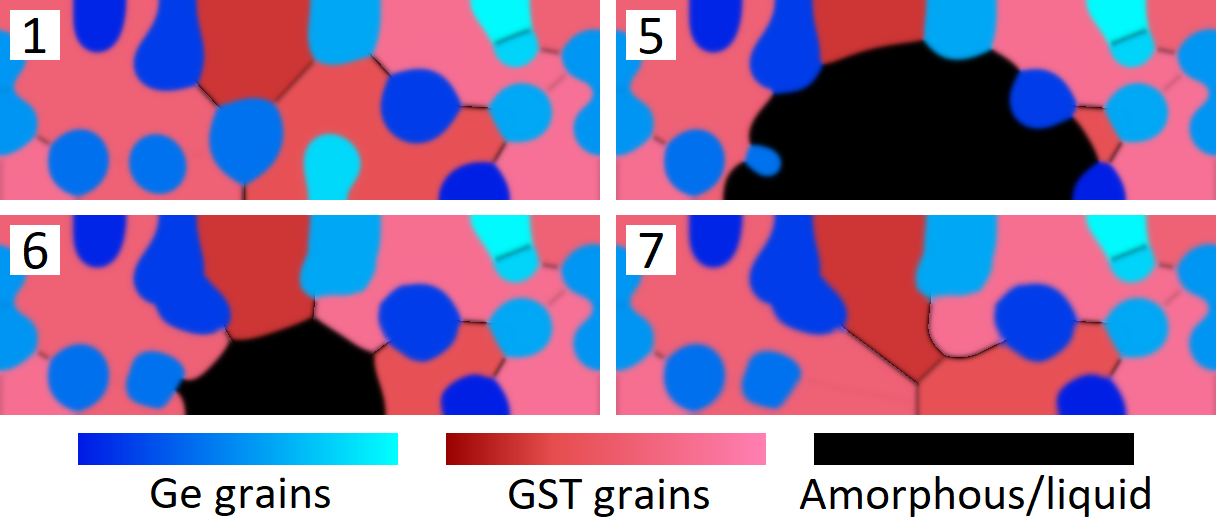}
	\caption{Microstructure evolution during a SET pulse: from the largest dome after initial heating (5) to a fully crystallized layer (7).}
	\label{fig_set_structure}
\end{figure}

Starting from the same crystallized layer, the evolution of the microstructure during a SET operation (pulse pattern of Fig. \ref{fig_pulses}) is shown in Fig. \ref{fig_set_structure}. Similar melting occurs but the progressive decrease of imposed current and temperature in the structure enables full recrystallization.

\subsection{Electrical figures of merit}
\begin{figure}[!htbp]
	\centering
	\includegraphics[width=0.82\columnwidth]{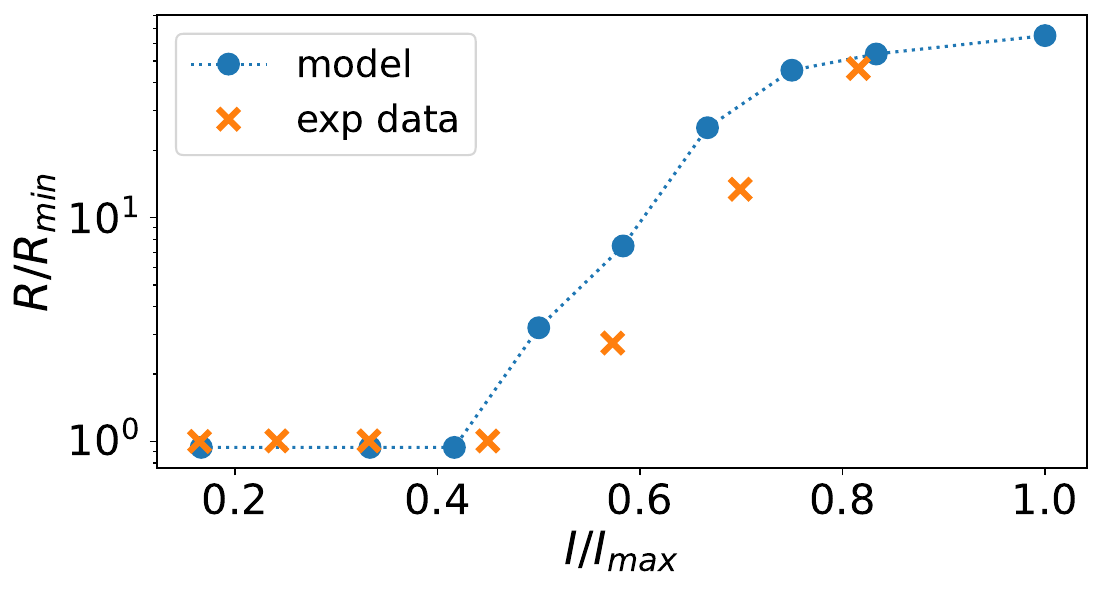}
	\caption{$R(I)$ simulation performed on the memory. Normalized electric current $I/I_{\max}$ and resistance $R/R_{\min}$.}
	\label{fig_R(I)}
\end{figure}

\begin{figure}[!htbp]
	\centering
	\includegraphics[width=0.82\columnwidth]{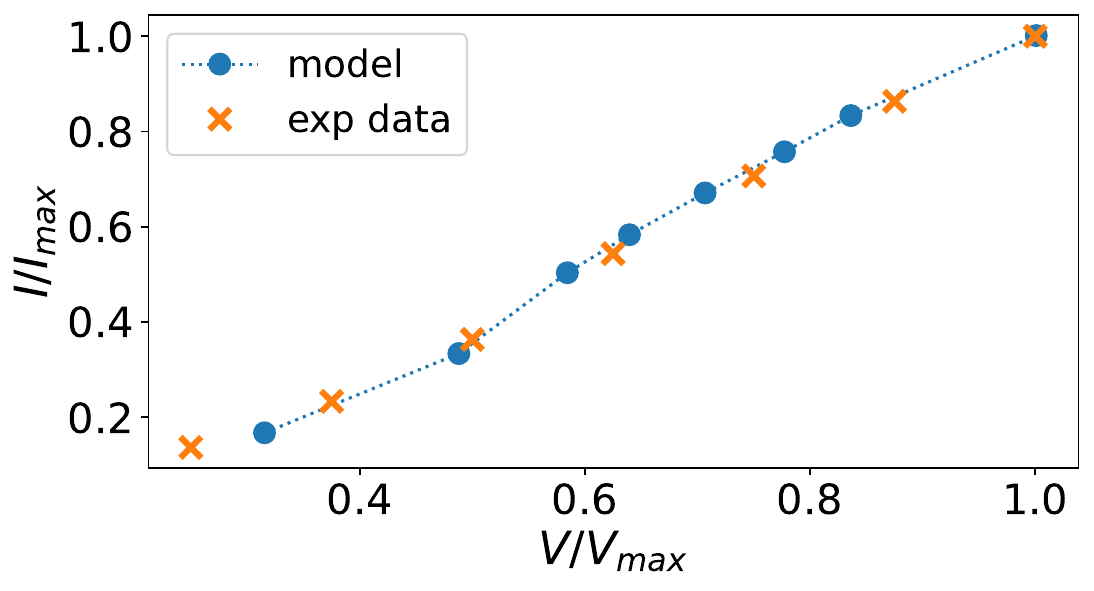}
	\caption{$I(V)$ simulation performed on the memory. Normalized voltage $V/V_{\max}$ and electric current $I/I_{\max}$.}
	\label{fig_I(V)}
\end{figure}

Simulation results have also been compared to electrical characterizations. Two figures of merit are reproduced: a
$R(I)$ plot in Fig. \ref{fig_R(I)} (RESET operations at various programming currents $I$ followed by the reading of $R$, the memory cell resistance)
and a $I(V)$ plot in Fig. \ref{fig_I(V)} (during sufficiently long RESET pulses at various fixed currents, the voltage V across the cell is read when the microstructure stops evolving and $V$ is stabilized).
In Fig. \ref{fig_R(I)}, the model qualitatively reproduces experimental measurements.
At a certain current, RESET resistance starts to increase (dome amorphization) and saturates at higher currents as the dome radius reaches the GGST layer thickness.
The resistances at low and high currents correspond to the electrical properties of the fully crystallized layer and of the amorphous phase, respectively.
To reproduce the gradual increase of resistance observed in experimental data, a Poole-Frenkel conduction in the amorphous phase and the high electrical conductivity of the liquid phase are essential.
The former reduces the resistivity of small amorphous domes, while the latter enables more progressive dome sizes by removing most of the Joule heating in the phase change layer as soon as the material melts.
In Fig. \ref{fig_I(V)}, the agreement is excellent.


\subsection{Impact of the initial microstructure}
\begin{figure}[!htbp]
	\centering
	\begin{subfigure}[b]{\columnwidth}
		\hspace{0.67cm}
		\includegraphics[width=0.82\columnwidth]{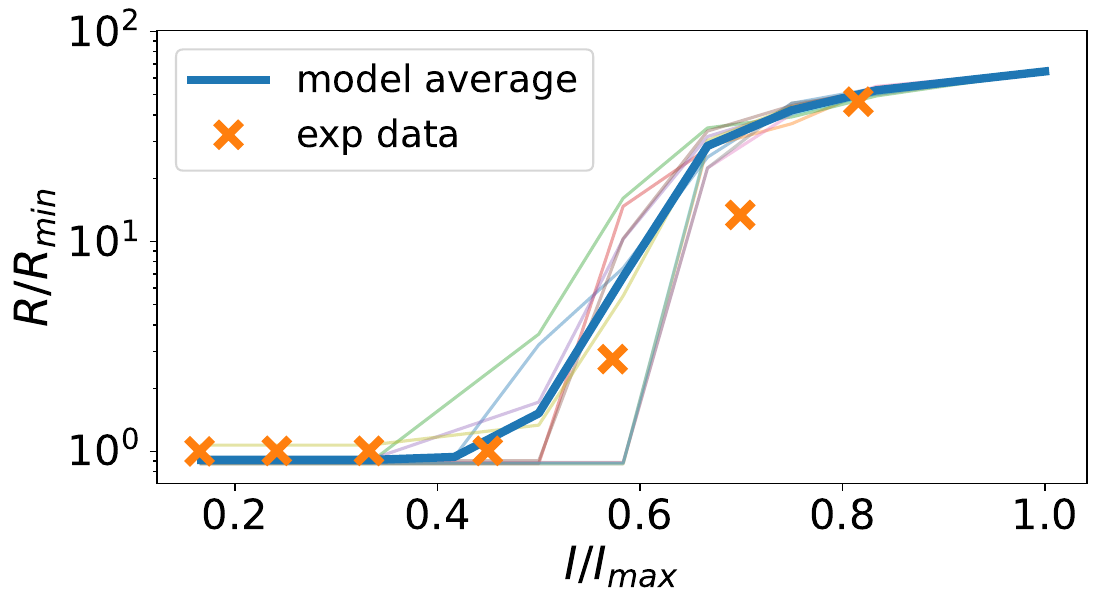}
	\end{subfigure}
	\begin{subfigure}[b]{\columnwidth}
		\hspace{0.67cm}
		\includegraphics[width=0.82\columnwidth]{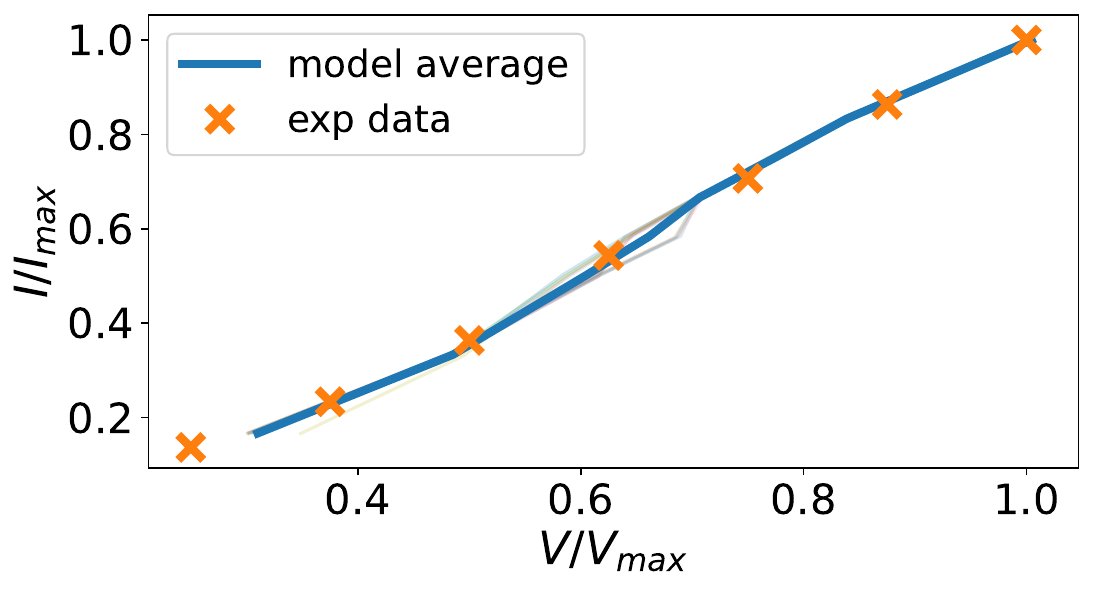}
	\end{subfigure}
\caption{Average of $R(I)$ and $I(V)$ simulations performed on various initial microstructures.
Individual simulations are displayed in the background with thinner lines.
Same normalization as in Fig. \ref{fig_R(I)} and \ref{fig_I(V)}.}
\label{fig_stat}
\end{figure}

Finally, the impact of the initial microstructure has been studied.
A total of 10 microstructures were generated randomly by multiple annealing simulations, and both $R(I)$ and $I(V)$ simulations were performed with each of them.
In Fig. \ref{fig_stat}, the curves associated with the different microstructures are averaged and compared to experimental data.
Individual curves are also displayed with thinner lines to visualize the variability of the results.
Although the individual curves are scattered, particularly for $R(I)$ simulations, the average remains in good agreement with experimental data.

\section{Conclusion}
In order to model the evolution of the GGST microstructure during memory operations, a fully self-consistent coupling of the MPFM with electro-thermal equations is presented.
Thermal and electrical parameters have been calibrated with literature data and in-house material characterization.
Even if an extended calibration is still needed to get more quantitative results, SET and RESET operation simulations demonstrate the ability of this model to qualitatively reproduce the GGST amorphization and crystallization depending on the programming conditions. Also, electrical figures of merit are in good agreement with experimental data.

This model provides an excellent framework to investigate other programming conditions, material changes and even emerging PCM applications such as analog programming, a growing subject linked to neuromorphic in-memory computing \cite{sebastian2019}.


\end{document}